\documentclass[10pt,british,tightenlines,eqsecnum,floats,aps,amsmath,amssymb,nofootinbib,superscriptaddress,prd,showpacs,showkeys]{revtex4}
\usepackage[latin9]{inputenc}
\usepackage{color}
\usepackage{babel}
\usepackage{amsmath}
\usepackage{amssymb}
\usepackage{graphicx}
\usepackage{esint}
\usepackage[unicode=true,pdfusetitle,
 bookmarks=true,bookmarksnumbered=false,bookmarksopen=false,
 breaklinks=false,pdfborder={0 0 1},backref=false,colorlinks=false]
 {hyperref}

\makeatletter
\@ifundefined{textcolor}{}
{%
 \definecolor{BLACK}{gray}{0}
 \definecolor{WHITE}{gray}{1}
 \definecolor{RED}{rgb}{1,0,0}
 \definecolor{GREEN}{rgb}{0,1,0}
 \definecolor{BLUE}{rgb}{0,0,1}
 \definecolor{CYAN}{cmyk}{1,0,0,0}
 \definecolor{MAGENTA}{cmyk}{0,1,0,0}
 \definecolor{YELLOW}{cmyk}{0,0,1,0}
}

\usepackage{color}\usepackage{euscript}\usepackage{epsfig}
\usepackage{amsfonts}\usepackage{fancyhdr}

\@ifundefined{textcolor}{}{%
 \definecolor{BLACK}{gray}{0}
 \definecolor{WHITE}{gray}{1}
 \definecolor{RED}{rgb}{1,0,0}
 \definecolor{GREEN}{rgb}{0,1,0}
 \definecolor{BLUE}{rgb}{0,0,1}
 \definecolor{CYAN}{cmyk}{1,0,0,0}
 \definecolor{MAGENTA}{cmyk}{0,1,0,0}
 \definecolor{YELLOW}{cmyk}{0,0,1,0}
}

\makeatother

\begin{document}

\title{\bf The special Two zeros texture based on $A_4$ symmetry and perturbation method}

\author{N. Razzaghi}
\email{n.razzaghi@qiau.ac.ir}

\affiliation{Department of Physics, Qazvin Branch, Islamic Azad
University, Qazvin, Iran}













\begin{abstract}
The current study aimed to investigate the special case of two
zeros in a Majorana neutrino mass matrix based on $A_4$ symmetry,
where charged lepton mass matrix is diagonal. The texture
$M_\nu^{S_7}$ with $(\mu, \mu)$ and $(\tau,\tau)$ vanishing
element of mass matrix has magic and $\mu-\tau$ symmetry, with a
tribimaximal form of the mixing matrix which leads to
$\theta_{13}=0$ that it is not consistent with experimental data
and does not seem to be allowed. Since $\theta_{13}$ is small
compared to other neutrino mixing angles, we show that
$\theta_{13}$, and $\delta$ could be obtained by using a complex
symmetrical perturbation in the mass basis and also could be shown
that $\delta m^2\equiv~m_2^2-m^2_1\neq0$ affecting the atmospheric
mixing angle.

We find that for the complex perturbation mass matrix, only the
results of the case I, $\Delta<0$ and $\textit{Re}(\alpha)<0$, are
consistent with experimental data. Furthermore, the allowed range
of our parameter space and complex element of perturbation are
found which led to finding the deviation of $\theta_{23}$ from
$45^\circ$ where this deviation is in line with the experimental
data which indicate the accuracy of our model and its results. Our
prediction is inverted mass ordering in the Case I. The results of
the case II,  $\Delta>0$ and $\textit{Re}(\alpha)>0$, are ruled
out.
\end{abstract}

\keywords{Two-zero texture; $A_4$ symmetry; Tribimaximal mixing
matrix, perturbation method, CP violation phases}

\date{25 November 2022}


\maketitle

\section{Introduction}
\label{int}

In the last two decades, neutrino experiments have illustrated
that neutrinos oscillate and are massive. Nevertheless, according
to the standard parametrization, the unitary lepton mixing matrix,
which connects the neutrino mass eigenstates to flavor
eigenstates, is given by \cite{mixing1, mixing2, mixing3}
\begin{equation}\label{emixing}
U_{PMNS}=\left(\begin{array}{ccc}c_{12}c_{13} & s_{12}c_{13} & s_{13}e^{-i\delta}\\
-s_{12}c_{23}-c_{12}s_{23}s_{13}e^{i\delta} &
c_{12}c_{23}-s_{12}s_{23}s_{13}e^{i\delta} &
s_{23}c_{13}\\s_{12}s_{23}-c_{12}c_{23}s_{13}e^{i\delta} &
-c_{12}s_{23}-s_{12}c_{23}s_{13}e^{i\delta} &
c_{23}c_{13}\end{array}\right),
\end{equation}
where $c_{ij}\equiv\cos\theta_{ij}\text{ and
}s_{ij}\equiv\sin\theta_{ij}$ (for $i,j=(1,2), (1,3) \text{ and }
(2,3) $); $ \delta $ is called the Dirac phase, analogous to the
CKM phase.

Finally, consequences of the neutrino experiments such as T2K
\cite{26, 27}, RENO \cite{28}, DOUBLE-CHOOZ \cite{29}, and
DAYA-BAY \cite{30, 31} have indicated that there are
 a nonzero mixing angle $\theta_{13}$ which is small compared
to the other two mixing ones and a possible nonzero Dirac
CP-violation phase $\delta_{CP}$. Therefore, the Tribimaximal
(TBM) mixing matrix is rejected \cite{32, 33}. The TBM mixing
matrix is \cite{TBM}:
\begin{equation}\label{etbm}
U_{TBM} =\left(\begin{array}{ccc}-\sqrt{\frac{2}{3}} & \frac{1}{\sqrt{3}} & 0\\
\frac{1}{\sqrt{6}} & \frac{1}{\sqrt{3}} &
-\frac{1}{\sqrt{2}}\\\frac{1}{\sqrt{6}} & \frac{1}{\sqrt{3}} &
\frac{1}{\sqrt{2}}\end{array}\right),
\end{equation}
where, regardless of the model, the mixing angles are:
$\theta_{12}\approx {35.26}^\circ$, $\theta_{13}\approx 0$, and
$\theta_{23}\approx 45^\circ$ \cite{25}.

Prior to this observation, models leading to the TBM mixing
matrix, widely studied \cite{256, 259}. Therefore, in order to
produce $\theta_{13}\neq 0$ starting from an initial TBM mixing
matrix, different approaches have been adopted \cite{2591}. One of
the successful phenomenological neutrino mass models with flavor
symmetry, which is an appropriate framework for understanding the
family structure of charged-lepton and of neutrino mass matrices
\cite{ma, ma1}, is illustrated by the group $A_4$ \cite{16, 17,
18, 19, 20, 21, 22, 23, 24}. The $A_4$ is a symmetry group of the
tetrahedron, was initially presented to illustrate a TBM mixing
matrix \cite{18}. Although, the primary initial objective of the
$A_4$ models was illustrating a TBM mixing matrix \cite{18}, many
efforts, e.g., \cite{16, 17}, \cite{19, 20, 21, 22, 23, 24},
\cite{38}, \cite{39}, \cite{43, 44, 45, 46}, have been made to set
up a model capable of describing the non-TBM mixing matrix
phenomenology.

The present $3\sigma$ global fits for the existing and known
neutrino oscillation parameters\cite{exp}:

\begin{eqnarray}\label{exp}
\delta m^{2}[10^{-5}eV^{2}]&=&(6.94-8.14),\nonumber\\
|\Delta m^{2}|[10^{-3}eV^{2}]&=&(2.47-2.63)-(2.37-2.53),\nonumber\\
\sin^{2}\theta_{12}&=&(0.271-0.369),\nonumber\\
\sin^{2}\theta_{23}&=&(0.434-0.610)-(0.433-0.608),\nonumber\\
\sin^{2}\theta_{13}&=&(0.02000-0.02405)-(0.02018-0.02424),\nonumber\\
\delta&=&(128^\circ-359^\circ)-(200^\circ-353^\circ),
\end{eqnarray}

multiple sets of allowed ranges are stated, and the left and the
right columns correspond to normal hierarchy and inverted
hierarchy, respectively. $\delta m^2\equiv m_2^2-m_1^2$ and
$\Delta m^2\equiv m_3^2-m_1^2 $.

Despite the prevailing information about neutrino oscillation
parameters (\ref{exp}), the mass and mixing problem in the lepton
sector is still conceived as a fundamental problem.

In the current work, we mainly focused on the neutrinos based on
especial case of two-zero textures with $A_4$ symmetry which is
here called $M_\nu^{S_7}$ \cite{200}. In \cite{200}, we study all
seven possible two-zero textures with $A_4$ symmetry, among which
only two textures, the texture with $(e, e)$ and $(e,\mu)$
vanishing element of mass matrix and its permutation symmetry, are
consistent with the experimental data in the non-perturbation
method.

In this paper, we intend to consider $M_\nu^{S_7}$ in perturbation
method to generate I) non-zero $\theta_{13}$, II) CP violation
phase $\delta$ and III) deviations of $\theta_{23}$ from
$\frac{\pi}{4}$. However, the discovery of the $\theta_{13}$,
whose smallness (in comparison to other mixing angles) signifies
modifying the neutrino mixing matrix by means of a small
perturbation about the basic TBM mixing matrix. By employing
different methods in a wide range of contexts, a lot of attempts
have been made to generate some of the neutrino parameters in
perturbation theory \cite{pur39}.

In the basis where the charged-lepton mass matrix is diagonal, a
particular application of $A_4$ is given by \cite{ma1}:
\begin{equation}\label{ema}
\cal{M}_\nu =\left(\begin{array}{ccc}a+\frac{2d}{3} & b-\frac{d}{3} & c-\frac{d}{3}\\
b-\frac{d}{3} & c+\frac{2d}{3} & a-\frac{d}{3}\\c-\frac{d}{3} &
a-\frac{d}{3}& b+\frac{2d}{3}\end{array}\right),
\end{equation}
which has also magic symmetry\footnote{Magic symmetry is a
symmetry in which the sum of elements in either any rows or any
columns of the neutrino mass matrix is identical \cite{magic}.}.

 Various phenomenological textures, specifically
texture zeros \cite{kumar,z8, z9, z10, z11, z12, z13, z14, z15,
permutation}, have been investigated in both flavor and non-flavor
bases. Such texture zeros not only causes to reduce the number of
free parameters of neutrino mass matrix, but also contributes to
establishing several simple and interesting relations between
mixing angles. Therefore, in the current research, this allowed us
to explore the effects of special case of two zero textures on
$\cal{M}_\nu $ given by~(\ref{ema}).

Moreover, assuming the Majorana nature of neutrinos, the present
study strove to investigate the phenomenological implications of
special case of two-zero textures of neutrino mass matrix together
with $A_4$ symmetry, based on a global fit of current neutrino
oscillation data \cite{exp}. The special case of two-zero texture
is $M_\nu^{S_7}$ with $(\mu,\mu)=0$, and $(\tau,\tau)=0$ which can
expose the impressive phenomenological features of a defined
Majorana neutrino mass matrix.

The organization of the paper is as follows. In Sec. II, the
methodology is elaborated in two subsections. In subsection A,
$M_\nu^{S_7}$ in the flavor bases was reconstructed as unperturbed
Neutrino Mass Matrix, and also unperturbed neutrino mass matrix
was obtained in the mass bases. In subsection B, the perturbed
neutrino mass matrix was presented as a complex symmetric
non-Hermitian matrix in the mass basis. The first-order of
neutrino mass correction and the third mass eigenstate were
obtained in the mass basis to first order corrections. Then
$|\nu_3\rangle_{mass}$ in the flavor bases was rewritten and
thereby $\sin^2{\theta_{13}}$, CP violation phase $\delta$ and
$\tan^2{\theta_{23}}$ were obtained. In Sec. III, the results are
compared with those of the experimental data in two different
cases. In each case, the complex elements of perturbation,
$\alpha$ and $\beta$, are illustrated onto the allowed region of
the parameter space and their allowed region were found. In the
case I, the allowed region of $\alpha$ and $\beta$ was acceptable;
therefore, the magnitude of $\theta_{23}$ was obtained which was
consistent with the experimental data and demonstrated the
accuracy of our work. In the case II, the obtained region of
$\alpha$ and $\beta$ was not acceptable; therefore it was ruled
out. In Sec. IV, the conclusions are provided.

\section{Methodology}
\subsection{The Unperturbed Neutrino Mass Matrix}

 Assuming the Majorana nature of neutrinos, the mass
matrix $\cal{M}_\nu $ is a complex symmetric matrix in
Eq.\,~(\ref{ema}). We examine in~(\ref{200}) the analysis of
two-zero texture for the Majorana neutrino mass matrix based on
$A_4$ symmetry, $\cal{M}_\nu$ in Eq.\,(\ref{ema}) restricts the
number of probabally viable cases to seven. It is found that the
texture with $(e, e)$ and $(e, \mu)$ vanishing elements of
$\cal{M}_\nu $ in~(\ref{ema}) and its permutation symmetry, are
consistent with the experimental data in the non-perturbation
method. The texture with $(e, \tau)=(\mu, \mu)=0$ and the texture
with $(e, \mu)=(\mu, \mu)=0$ and their permutation symmetry, are
not consistent with the experimental data at all in~(\ref{200}).
The seventh probably viable case of two-zero texture of
$\cal{M}_\nu$ in Eq.\,(\ref{ema}) with $(\mu, \mu)=(\tau, \tau)=0$
is extremely interesting. We call it $M_\nu^{S_7}$ in~(\ref{200})
and is given by;
\begin{equation}\label{emm7}\vspace{.2cm}
M_\nu^{S_7} =\left(\begin{array}{ccc}a+\frac{2}{3}d & -d &  -d\\
-d &  0 &a-\frac{d}{3}\\ -d &a-\frac{d}{3}& 0
\end{array}\right).
\end{equation}

This mass matrix is a magic matrix, which obviously has $\mu-\tau$
symmetry. Consequently, it can lead to TBM mixing matrix~ in
Eq.\,(\ref{etbm}) with $\theta_{13}=0$. Therefore, initially seems
that $M_\nu^{S_7}$~in Eq.\,(\ref{emm7}) is not allowed texture,
but we consider it in perturbation method.

A straightforward diagonalization procedure yields
$U_{TBM}^{T}M_\nu^{S_7}U_{TBM}=M_{diag}^{S_7}$, where
\begin{eqnarray}\label{emm1}
m_1&=&a+\frac{5}{3}d,\nonumber\\
m_2&=&a-\frac{4}{3}d ,\nonumber\\
m_3&=&-a+\frac{1}{3}d,
\end{eqnarray}
the mass eigenvalues can be complex, they can be presented
positive and real by phase transformation, as $diag=(1,
e^{i\rho},e^{i\sigma})$, which $\rho$ and $\sigma$ are Majorana
phases where neutrino oscillations are independent of them.

We reconstruct $M_\nu^{S_7}$ in Eq.\,(\ref{emm7}) by using
$M_\nu=U^{*}_{TBM}M_{diag}U^\dagger_{TBM}$, where $M_\nu$ is a
magic neutrino mass matrix with $\mu-\tau$ symmetry. In this
reconstruction, we define new parameters, as
\begin{eqnarray}\label{emm2}
m&\equiv&\frac{\sum m_i}{3}=\frac{(m_1+m_2+m_3)}{3},\nonumber\\
\Delta_{32}&\equiv&(m_3-m_2) ,\nonumber\\
\Delta_{31}&\equiv&(m_3-m_1).
\end{eqnarray}

Also because of; the reported experimental results which have
shown $\Delta m^2_{21}$ is tiny and greater than zero \cite{exp},
we approximate $\Delta_{31}\simeq\Delta_{32}\equiv\Delta$.
Therefore, the unperturbed mass matrix, the reconstructed
$M_\nu^{S_7}$, in the flavor basis\footnote{We work in a basis
where the charged lepton mass matrix is diagonal, and thereby, the
lepton mixing is extracted from the neutrino mass matrix.} is;

\begin{equation}\label{eunp}
M^{0}_{f\nu} \simeq \left(\begin{array}{ccc}m-\frac{\Delta}{3} & 0 & 0\\
0 & m+\frac{\Delta}{6} &-\frac{\Delta}{2}\\0 &-\frac{\Delta}{2}&
m+\frac{\Delta}{6}
\end{array}\right).
\end{equation}
At this level, the unperturbed mass matrix $M^{0}_{f\nu}$ in
Eq.\,(\ref{eunp}) has $\mu-\tau$ symmetry but is no longer magic.
The mass spectrum of $M^{(0)}_{f\nu}$ is
\begin{eqnarray}\label{emm2}\vspace{.2cm}
m^{(0)}_1 = m^{(0)}_2 =m-\frac{\Delta}{3} ,
~~~~\text{and}~~~m^{(0)}_{3}=m+\frac{2\Delta}{3}.
\end{eqnarray}
Here $m^{(0)}_1$, $m^{(0)}_2$ and $m^{(0)}_3$ are real and
positive and $m^{(0)}_1$, and $m^{(0)}_2$ are the same. Hence, the
unperturbed mass matrix in the mass basis is;
\begin{equation}\label{eunpm}
M^{0}_{mass} \simeq \left(\begin{array}{ccc}m-\frac{\Delta}{3} & 0 & 0\\
0 & m-\frac{\Delta}{3} &0\\0 &0& m+\frac{2\Delta}{3}
\end{array}\right).
\end{equation}
In the mass basis the eigenstates of the unperturbed neutrino mass
matrix $M^{0}_{mass}$ in Eq.\,(\ref{eunpm})are as follows:

\begin{equation} \vspace{.2cm}\label{em0}
|\nu^{(0)}_{1}\rangle=\left(\begin{array}{ccc}1\\
0\\0
\end{array}\right), ~~~~|\nu^{(0)}_{2}\rangle=\left(\begin{array}{ccc}0\\
1\\0\end{array}\right), ~~~~~|\nu^{(0)}_{3}\rangle=\left(\begin{array}{ccc}0 \\
0\\1\end{array}\right),
\end{equation}
in which the first two mass eigenstates are degenerate and the
columns of $U_{TBM}$ in Eq.\,(\ref{etbm}) are the unperturbed
flavour eigenstates.

We should mention that, up to now, the shortcomings are: (i) the
absence of $m^{(0)}_1$, and $m^{(0)}_2$  splitting, (ii) the
ordering of neutrino masses which is unknown and (iii) the mixing
matrix which is still $U_{TBM}$. Thus, the main objective is
obtaining the splitting of $m^{(0)}_1$, and $m^{(0)}_2$ by means
of a mass perturbation, from which $\theta_{13}\neq 0 $ and CP
violation are also derived. Moreover, CP violation conditions are
necessarily mandate in that $\mu-\tau$ symmetry should be broken.
An interesting question is: After breaking the $\mu-\tau$
symmetry, will $\theta_{23}=45^{\circ}$ remain valid or not?

\subsection{The perturbed Neutrino Mass Matrix and Perturbation}

In our work, the minimal symmetric perturbation neutrino mass
matrix in the mass basis can be written as~\footnote{It is a
general form of perturbation mass matrix for nonmagic unperturbed
neutrino mass matrix with $\mu-\tau$ symmetry where the first two
mass eigenvalues are degenerate as we work in \cite{meFL3}. };

\begin{equation}\label{epm}
M^{\prime}_{mass} \simeq \Delta \left(\begin{array}{ccc}0 & 0 & \beta\\
0 & \alpha &0\\\beta &0& 0
\end{array}\right).
\end{equation}

The dimensionless perturbation elements i.e., $\alpha ,\beta$ can
be real or complex and should be small compared to the elements of
unperturbed neutrino mass matrix $M^{0}_{mass}$~in
Eq.\,(\ref{eunpm}) for a valid perturbation theory.

If $M^{\prime}_{mass}$ is a complex symmetric non-Hermitian matrix
given in Eq.\,(\ref{epm}), therefore we have to consider
$(M^{0}_{mass}+M^{\prime}_{mass})^\dagger(M^{0}_{mass}+M^{\prime}_{mass})$
where we drop a term, which is ${\cal O} (\alpha^2,~\beta^2)$.
${M^{0}_{mass}}^{\dagger} M^{0}_{mass}$ is the unperturbed
Hermitian term , and its eigenstates are the same as those of
$M^{0}_{mass}$~in Eq.\,(\ref{em0}) and its eigenvalues are
$(m^{(0)}_1)^2$, $(m^{(0)}_2)^2$ and $(m^{(0)}_3)^2$, therefore
the perturbation term is
$M^p_{mass}={M^{0^{\dagger}}_{mass}}M^{\prime}_{mass}+{M^{\prime^{\dagger}}_{mass}}M^{0}_{mass}$
and the perturbation matrix is

\begin{equation}\label{epm1}
M^{p}_{mass} \simeq \Delta~\left(\begin{array}{ccc}0 & 0 & \textit{Re}(\beta)(2m+\frac{\Delta}{3})+\textit{Im}(\beta)(-\Delta)\\
0 &  \textit{Re}(\alpha)(2m-\frac{2\Delta}{3})
&0\\\textit{Re}(\beta)(2m+\frac{\Delta}{3})+\textit{Im}(\beta)(\Delta)
&0& 0
\end{array}\right).
\end{equation}
The first-order corrections to the neutrino masses, in the mass
basis~Eq.\,(\ref{em0}), are obtained from
$m_i^{(1)}\delta_{ij}=\langle\nu_i^{(0)}|{M^{0^{\dagger}}_{mass}}M^{\prime}_{mass}+{M^{\prime^{\dagger}}_{mass}}M^{0}_{mass}|\nu_j^{(0)}\rangle$.
 Therefore, by using~Eq.\,(\ref{emm2}) and the first-order of neutrino mass
correction, we have
\begin{eqnarray}\label{emm3}
m_1^2&=&(m^{(0)}_1)^2 ,\nonumber\\
m_2^2&=&(m^{(0)}_2)^2 +\textit{Re}(\alpha)(2m-\frac{2}{3}\Delta)\Delta,\nonumber\\
m_3^2&=&(m^{(0)}_3)^2.
\end{eqnarray}
The mass correction arises from this order corrections with
$m_2^{(1)}\neq0$, and $m_1^{(1)}=m_3^{(1)}=0$. Therefore, the
splitting of $m_1$, and $m_2$ is equal to $m_2^{(1)}$. Neutrino
experimental data have so far definitely confirmed that $\delta
m^{2}=m_2^2-m_1^2>0$. Therefore, by using ~Eq.\,(\ref{emm2})
$m_2^{(1)}=\textit{Re}\alpha(2~m^{(0)}_2)\Delta$ must be positive.

From equations~in ~Eq.\,(\ref{emm3}), we could obtain the ratio of
two neutrino mass-squared differences $R_\nu=\frac{\delta
m^2}{\Delta m^2}$, as
\begin{equation}\label{er}\vspace{.2cm}
R_\nu=\textit{Re}(\alpha)\frac{6m-2\Delta}{6m+\Delta},
\end{equation}
where $\delta m^2\equiv m_2^2-m_1^2$ and $\Delta m^2\equiv
m_3^2-m_1^2 $.

We reproduce the third mass eigenstate $|\nu_3>$, in the mass
basis to the first order corrections, by
\begin{equation}\label{enu3}
|\nu_3>_{mass}=|\nu^{(0)}_{3}\rangle+\frac{ <\nu^{(0)}_j|
M^{p}_{mass} |\nu^{(0)}_3> }{(m^{(0)}_3)^2 -
(m^{(0)}_j)^2}|\nu^{(0)}_{j}\rangle, \;\; (j \neq 3). \;
\end{equation}

Using Eq.\,(\ref{emm2}), Eq.\,(\ref{em0}) and~Eq.\,(\ref{epm1}),
we could obtain $|\nu_3>_{mass}$ as

\begin{equation}\label{enu33}
|\nu_3>_{mass}=\left(\begin{array}{ccc}\textit{Re}(\beta)-\frac{3\Delta}{6m+\Delta}\textit{Im}(\beta)\\
0\\1
\end{array}\right).
\end{equation}

Now, we rewrite $|\nu_3>_{mass}$ in~ Eq.\,(\ref{enu33}), in the
flavor basis, as follows

\begin{equation}\label{enu3f}
|\nu_3>_{flavor}=\left(\begin{array}{ccc}\sqrt{\frac{2}{3}}\left(\frac{-3\Delta\textit{Im}(\beta)}{6m+\Delta}+\textit{Re}(\beta)\right)\\
\frac{1}{\sqrt{2}}-\frac{\frac{-3\Delta\textit{Im}(\beta)}{6m+\Delta}+\textit{Re}(\beta)}{\sqrt{6}}\\-\frac{1}{\sqrt{2}}-\frac{\frac{-3\Delta\textit{Im}(\beta)}{6m+\Delta}+\textit{Re}(\beta)}{\sqrt{6}}
\end{array}\right),
\end{equation}

and then could obtain nonzero values for both $\sin\theta_{13}$ ,
$\delta$, and the deviation of $\theta_{23}$ from $45^{\circ}$ by
comparing $|\nu_3>_{flavor}$~in~ Eq.\,(\ref{enu3f}) with the third
column of lepton mixing matrix ~in~ Eq.\,(\ref{emixing}).
Therefore,

\begin{eqnarray}\label{es13}
\sin^2\theta_{13}&=&|U_{13}|^2= \frac{2}{3}(\textit{Re}(\beta))^2+\frac{6\Delta^2}{(6m+\Delta)^2}(\textit{Im}(\beta))^2,\nonumber\\
\tan\delta&=&\frac{(-3\Delta)\textit{Im}(\beta)}{(6m+\Delta)\textit{Re}(\beta)},\nonumber\\
\tan\theta^2_{23}&=&\frac{|U_{23}|^2}{|U_{33}|^2}=
1-\frac{4\sqrt{3}\textit{Re}(\beta)}{3+(\frac{3\Delta\textit{Im}(\beta)}{6m+\Delta})^2+2\sqrt{3}\textit{Re}(\beta)+(\textit{Re}(\beta))^2}.
\end{eqnarray}

Up to this point, having used perturbation method, we could obtain
I)the splitting of the two first neutrino masses, II)the third
perturbed mass eigenstate, in the mass basis~Eq.\,(\ref{enu33})
and also the flavor basis~Eq.\,(\ref{enu3f}), in the CP violation
case, therefor generating III)$\theta_{13}\neq0$, VI)$\delta$, and
V) the rate of $\theta_{23}$ deviation from $45^{\circ}$. In the
next section, by comparing the results of our work those of the
experimental data Eq.\,(\ref{exp}), we will show that
$M_\nu^{S_7}$~(\ref{emm7}) could be allowed texture.

\section{COMPARISON WITH EXPERIMENTAL DATA}

In this section the results of the current study are compared with
those of the experimental data. As it was mentioned in the
previous section, according to neutrino experimental data, we have
$\delta m^{2}=\textit{Re}(\alpha)(2~m^{(0)}_2)\Delta>0$.
Therefore, since $m^{(0)}_2$ in Eq.\,(\ref{emm2}) is real and
positive, we have two cases; I)$\Delta$ and $\textit{Re}(\alpha)$
to be negative, II)$\Delta$ and $\textit{Re}(\alpha)$ to be
positive.

\begin{figure}[th]
\includegraphics[width=8cm]{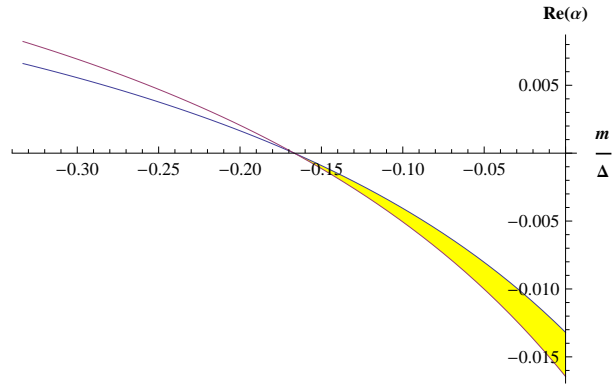}\caption{\label{fig.1} \small
(color online). The range of $\textit{Re}(\alpha)$ onto
$\frac{m}{\Delta}$ our parameter space, according to the
experimental data of $R_\nu$. The yellow (dark) area displays the
allowed region of $\textit{Re}(\alpha)$ within our model in the
case I.}
 \label{geometry}
\end{figure}

Initially, we consider the case I, when $\Delta<0$ and
$\textit{Re}(\alpha)<0$. The investigation of the case I includes
three steps, in the first step, we obtain the allowed range of
$\textit{Re}(\alpha)$, as it is shown in Figure \ref{fig.1}. We do
this by substituting the experimental range of $R_\nu$ ~in
Eq.\,(\ref{er}) and mapping $\textit{Re}(\alpha)$ onto our
parameter space of the case I,
$-\frac{1}{3}\leq\frac{m}{\Delta}\leq0$ ~\footnote{The allowed
range of our parameter space based on $m^{(0)}_2>0$}, according to
the experimental data of $R_\nu$.

As regards the case I, $\Delta$ and $\textit{Re}(\alpha) $ must be
negative; therefore, as it is shown in Figure \ref{fig.1} the
allowed range of $\textit{Re}(\alpha) $ is restricted to the below
of the horizontal axis, $(\frac{m}{\Delta})$ in the dark area
between two curves. Therefore, the allowed range of
$\textit{Re}(\alpha) $ is

\begin{equation}\label{eA}
\textit{Re}(\alpha)\approx~0\rightarrow~-(0.0131-0.0161).
\end{equation}

Likewise, as it is shown in Figure \ref{fig.1}, we could to
specify the allowed range of our parameter space,
$\frac{m}{\Delta}$ according to the allowed range of
$\textit{Re}(\alpha) $ as follows,

\begin{equation}\label{eA1}
\frac{m}{\Delta}\approx~(-0.167-0).
\end{equation}

\begin{figure}[th] \includegraphics[width=8cm]{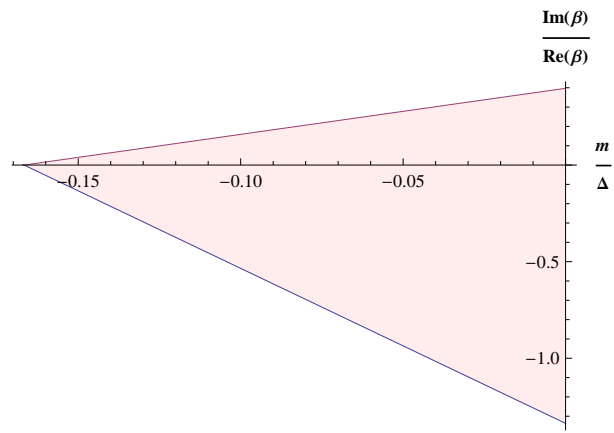}\caption{\label{fig.2} \small
(color online). The dark area is the allowed range of
$\frac{\textit{Im}(\beta)}{\textit{Re}(\beta)}$ onto
$\frac{m}{\Delta}$ our parameter space, according to the
experimental data of $\tan\delta$.}
 \label{geometry}
\end{figure}

In the second step , we obtain the allowed range of
$\frac{\textit{Im}(\beta)}{\textit{Re}(\beta)}$. As it is shown in
Figure \ref{fig.2}, we do this by using the equation of
$\tan\delta$~in Eq.\,(\ref{es13}) and mapping
$\frac{\textit{Im}(\beta)}{\textit{Re}(\beta)}$ onto the allowed
range of our parameter space~ Eq.\,(\ref{eA1}) according to the
experimental data of $\tan\delta$~ in Eq.\,(\ref{exp}).

Therefore, we obtain the allowed region of
$\frac{\textit{Im}(\beta)}{\textit{Re}(\beta)}$ onto our parameter
space which is consistent with the experimental data as follows
\begin{equation}\label{eBB}
\frac{\textit{Im}(\beta)}{\textit{Re}(\beta)}\approx((-1.31)-0.4),
\end{equation}
accordingly, can write
\begin{equation}\label{eBB1}
\textit{Im}(\beta)\approx((-1.31)-0.4){\textit{Re}(\beta)},
\end{equation}
therefore, in the second step, we obtain the ratio of
$\frac{\textit{Im}(\beta)}{\textit{Re}(\beta)}$.

\begin{figure}[th]
\includegraphics[width=8cm]{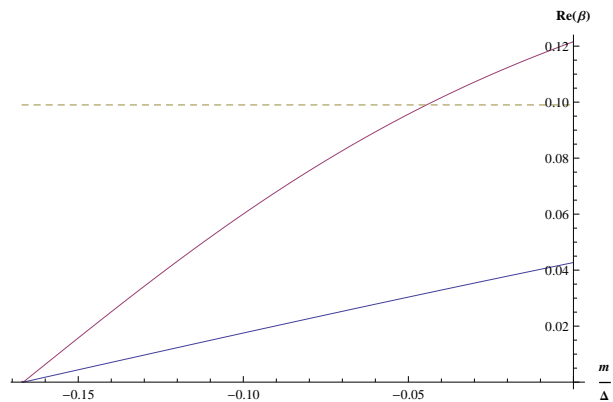}\caption{\label{fig.3} \small
(color online). Range of $\textit{Re}(\beta)$ onto
$\frac{m}{\Delta}$ our parameter space, according to the
experimental data of $\sin^2\theta_{13}$. The area bellow the line
$0.099$ displays the allowed region of $\textit{Re}(\beta)$ within
our model.}
 \label{geometry}
\end{figure}

In the third step, we obtain the allowed range of
$\textit{Re}(\beta)$, as it is shown in Figure \ref{fig.3}, and
subsequently the allowed range of ${\textit{Im}(\beta)}$ by
using~Eq.\,(\ref{eBB1}). We do this by using the equation of
$\sin^2\theta_{13}$ in Eq.\,(\ref{es13}) and mapping
$\textit{Re}(\beta)$ onto our parameter space~Eq.\,(\ref{eA1}),
according to the experimental data of $\sin^2\theta_{13}$~in
Eq.\,(\ref{exp}).

According to perturbation theory, the perturbation elements i.e.,
$\alpha$, and $\beta$ in Eq.\,(\ref{epm}) should be small compared
to the elements of unperturbed neutrino mass matrix $m$, and
$\Delta$ ~in Eq.\,(\ref{eunpm}). Accordingly, we only accept those
values of $\textit{Re}(\beta)$ that are less than
$0.1$.\footnote{We choose $\textit{Re}(\beta)<0.1$ because of the
experimental result for the sum of the three light neutrino masses
that has been reported by the Planck measurements of the cosmic
microwave background \cite{planck},~$\sum m_\nu<0.12eV $}
Therefore, the area bellow the line $0.099$ in figure \ref{fig.3}
displays the allowed region of $\textit{Re}(\beta)$ within our
model, as

\begin{equation}\label{eBB2}
\textit{Re}(\beta)\approx~0\rightarrow(0.043-0.099),
\end{equation}

then based on~Eq.\,(\ref{eBB1}), and Eq.\,(\ref{eBB1})
the allowed region of $\textit{Im}(\beta)$ is
\begin{equation}\label{eBB3}
\textit{Im}(\beta)\approx~0\rightarrow((-0.056)-0.04),
\end{equation}


Having taken these three steps, we obtain the allowed region of
our parameter space,$\frac{m}{\Delta}$~in Eq.\,(\ref{eA1}), and
dimensionless perturbation elements $\textit{Re}(\alpha)$,
$\textit{Re}(\beta)$, and $\textit{Im}(\beta)$~(\ref{eBB3}),
respectively in Eq.\,(\ref{eA}), Eq.\,(\ref{eBB2}), and
Eq.\,(\ref{eBB3}).

\begin{figure}[th]
\includegraphics[width=8cm]{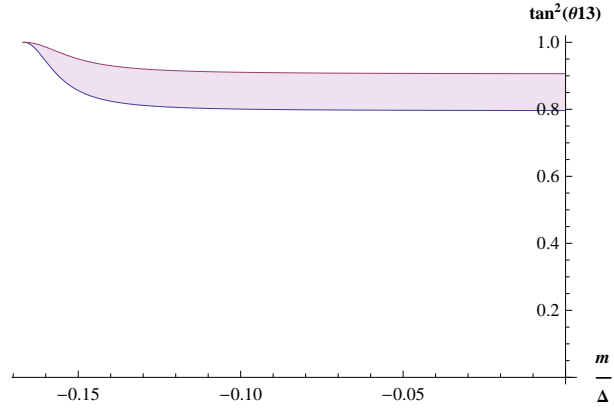}\caption{\label{fig.4} \small
(color online). Range of $\tan\theta^2_{23}$ onto
$\frac{m}{\Delta}$ parameter space, according to the allowed range
of perturbation elements. The dark area displays the allowed
region of $\tan\theta^2_{23}$ within our model which in complete
agreement with the experimental data ~in Eq.\,(\ref{exp}).}
 \label{geometry}
\end{figure}

Now we must examine the accuracy of the results of the case I by
determining the magnitude of $\tan\theta^2_{23}$~in
Eq.\,(\ref{es13}) by values of our parameter obtained in the
pervious three steps. We plot $\tan\theta^2_{23}$, based on the
allowed region of perturbation elements, onto parameter space as
it is shown in Figure \ref{fig.4}. Interestingly, the values
obtained for $\tan\theta^2_{23}$ corroborate those of with the
experimental data ~in Eq.\,(\ref{exp}). We find the allowed region
of $\theta_{23}$ in the case I as

\begin{equation}\label{eBB4}
\theta_{23}\approx(41.81^\circ-43.96^\circ)\rightarrow~45^\circ,
\end{equation}
which indicate the accuracy of the case I in our work .

Also according to the sign of $\Delta$ and $\textit{Re}(\beta)$ in
this case, the neutrino mass ordering is inverted.

\begin{figure}[th]
\includegraphics[width=8cm]{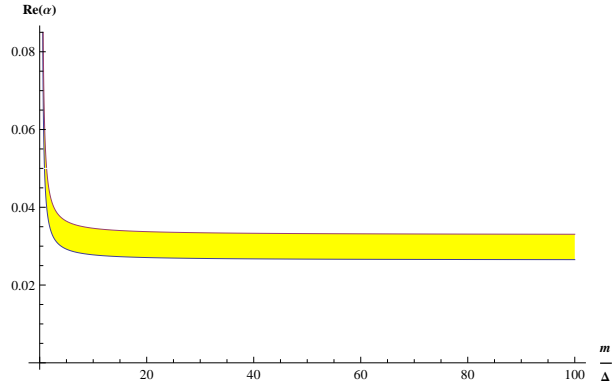}\caption{\label{fig.5} \small
(color online). The range of $\textit{Re}(\alpha)$ onto
$\frac{m}{\Delta}$, our parameter space of the case II, according
to the experimental data of $R_\nu$. The allowed range of
$\textit{Re}(\alpha) $ is restricted in the dark region between
two parallel horizontal lines in the case II.}
 \label{geometry}
\end{figure}

Now we consider the case II when  $\Delta$ and
$\textit{Re}(\alpha)$ are both positive. In this case, similar to
the case I, we first obtain the allowed range of
$\textit{Re}(\alpha)$, as it is shown in Figure \ref{fig.5}. We do
this by substituting the experimental range of $R_\nu$ ~in
Eq.\,(\ref{er}) and mapping $\textit{Re}(\alpha)$ onto our
parameter space of the case II, $\frac{m}{\Delta}\geq\frac{1}{3}$
/footnote{The allowed range of our parameter space based on
$m^{(0)}_2>0$}, according to the experimental data~in
Eq.\,(\ref{exp}).

Therefore in Case II, according to Figure \ref{fig.5}, the allowed
range of $\textit{Re}(\alpha) $ which is restricted in the dark
region is

\begin{equation}\label{eAA}
\textit{Re}(\alpha)\approx~(0.0281-0.04)\rightarrow\infty.
\end{equation}

\begin{figure}[th] \includegraphics[width=8cm]{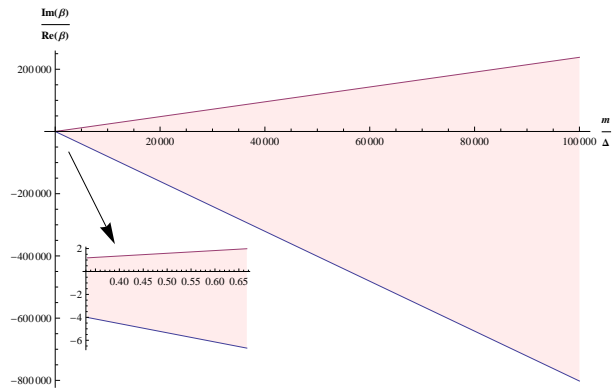}\caption{\label{fig.6} \small
(color online). The dark area is the allowed range of
$\frac{\textit{Im}(\beta)}{\textit{Re}(\beta)}$ in
$\frac{m}{\Delta}$ parameter space in the case II, according to
the experimental data of $\tan\delta$.}
 \label{geometry}
\end{figure}

In the next step, we find the allowed range of
$\frac{\textit{Im}(\beta)}{\textit{Re}(\beta)}$ in the case II as
shown in Figure \ref{fig.6}. Same as the case I, we do this by
using the equation of $\tan\delta$~in Eq.\,(\ref{es13}) and
mapping $\frac{\textit{Im}(\beta)}{\textit{Re}(\beta)}$, but onto
the allowed range of our parameter space in the case II, according
to the experimental data of $\tan\delta$~ in Eq.\,(\ref{exp}).

We find that the allowed region of
$\frac{\textit{Im}(\beta)}{\textit{Re}(\beta)}$ onto our parameter
space, in the case II, as
\begin{equation}\label{eBBB}
\frac{\textit{Im}(\beta)}{\textit{Re}(\beta)}\approx~-(4-\infty)~\text{and}~(1.2-\infty),
\end{equation}
therefore, according Eq.\,(\ref{eBBB})could write
\begin{equation}\label{eBBB1}
\textit{Im}(\beta)\approx~-(4-\infty)\textit{Re}(\beta)~\text{and}~(1.2-\infty)\textit{Re}(\beta).
\end{equation}

\begin{figure}[th]
\includegraphics[width=8cm]{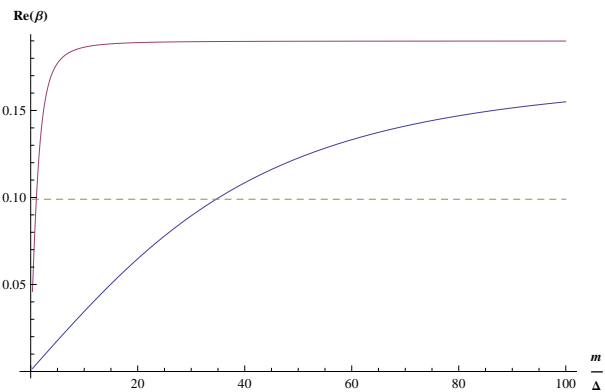}\caption{\label{fig.7} \small
(color online). The range of $\textit{Re}(\beta)$ onto
$\frac{m}{\Delta}$ our parameter space, according to the
experimental data of $\sin^2\theta_{13}$ when
$\textit{Im}(\beta)\approx~-(4-\infty)\textit{Re}(\beta)$. The
area bellow the horizontal dashed line displays the allowed region
of $\textit{Re}(\beta)$ in the case II within our model.}
 \label{geometry}
\end{figure}

\begin{figure}[th]
\includegraphics[width=8cm]{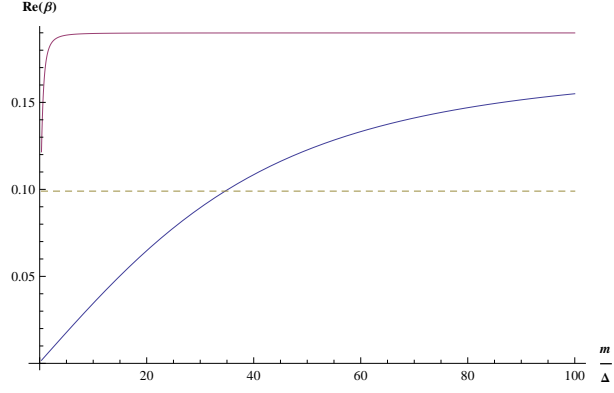}\caption{\label{fig.71} \small
(color online). The range of $\textit{Re}(\beta)$ onto
$\frac{m}{\Delta}$ our parameter space, according to the
experimental data of $\sin^2\theta_{13}$ when
$\textit{Im}(\beta)\approx~(1.2-\infty)\textit{Re}(\beta)$. The
area bellow the horizontal dashed line displays the allowed region
of $\textit{Re}(\beta)$ in the case II within our model.}
 \label{geometry}
\end{figure}

We obtain the allowed range of $\textit{Re}(\beta)$ in the two
different regions which are specified in Eq.\,(\ref{eBBB}), as
shown in the Figure \ref{fig.7}, and the Figure\ref{fig.71}
respectively. As the pervious case, we do this by inserting the
experimental data of $\sin^2\theta_{13}$ in the equation of
$\sin^2\theta_{13}$ in Eq.\,(\ref{es13}) and mapping
$\textit{Re}(\beta)$ onto our parameter space, in the case II.

\begin{figure}[th]
\includegraphics[width=8cm]{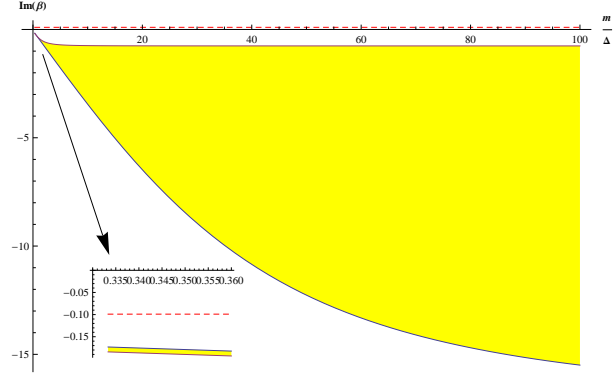}\caption{\label{fig.8} \small
(color online). The range of $\textit{Im}(\beta)$ onto
$\frac{m}{\Delta}$ our parameter space, according to the range of
$\textit{Re}(\beta)$ when
$\textit{Im}(\beta)\approx~-(4-\infty)\textit{Re}(\beta)$. The
dark area displays the obtained region of $\textit{Im}(\beta)$
which is not allowed within our model because is greater than
$|0.1|$ (the red dashed line).}
 \label{geometry}
\end{figure}

\begin{figure}[th]
\includegraphics[width=8cm]{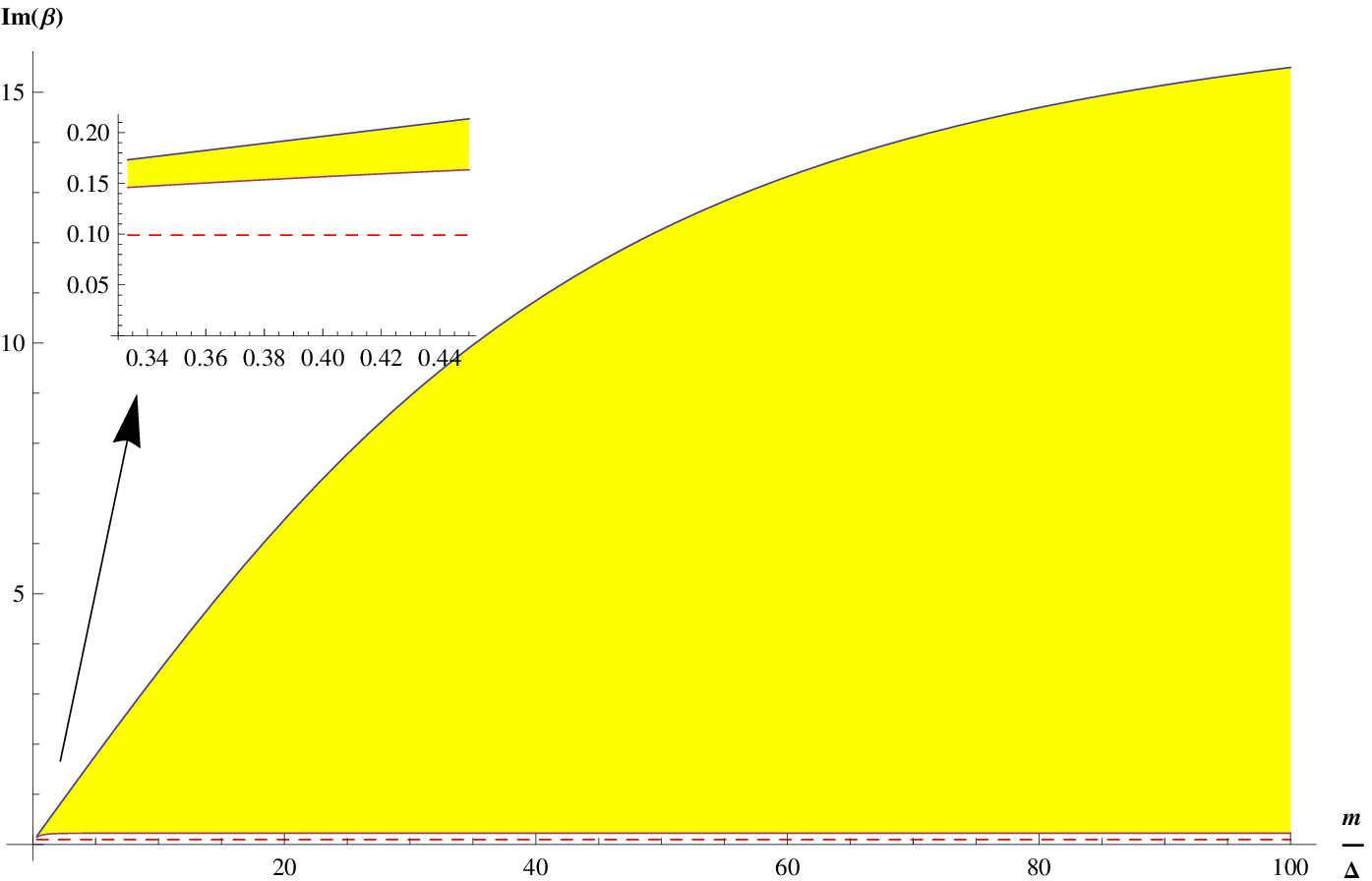}\caption{\label{fig.81} \small
(color online). The range of $\textit{Im}(\beta)$ onto
$\frac{m}{\Delta}$ our parameter space, according to the range of
$\textit{Re}(\beta)$ when
$\textit{Im}(\beta)\approx~(1.2-\infty)\textit{Re}(\beta)$. The
dark area displays the obtained region of $\textit{Im}(\beta)$
which is not allowed within our model because is greater than
$|0.1|$ (the red dashed line).}
 \label{geometry}
\end{figure}

As mentioned before, the allowed range of $\textit{Re}(\beta)$
must be bellow $0.099$; therefore, the area bellow the horizontal
dashed line in the Figures \ref{fig.8} and, the Figure
\ref{fig.81} displays the allowed range of $\textit{Re}(\beta)$ in
the case II according to the two ranges of specified in~
Eq.\,(\ref{eBBB1}).

Afterwards, we map ${\textit{Im}(\beta)}$ based on
Eq.\,(\ref{eBBB1}) ,in the two different range of
$\textit{Re}(\beta)$, as shown in the Figure \ref{fig.8} and, the
Figure \ref{fig.81} respectively. We find that in the case II the
obtained value of $\textit{Im}(\beta)$,
 is much greater than $|0.1|$, the red dashed line, which is not acceptable by the perturbation theory. Therefore, the obtained
results in the case II, with $\Delta>0$ and
$\textit{Re}(\alpha)>0$, are ruled out.

\section{Conclusion}
\label{concl}

We have studied the phenomenology of two-zero texture in the
Majorana neutrino mass matrix with $A_4$ symmetry where the
charged lepton mass matrix is diagonal. Therefore, there are seven
viable two-zero textures. The seven viable textures are broadly
categorized into two categories. To sum up, in general textures
which are consistent with the experimental data as $M_\nu^{S_1}$,
$M_\nu^{S_2}$, $M_\nu^{S_7}$ and textures which are not consistent
with the experimental data as $M_\nu^{S_3}$, $M_\nu^{S_5}$, and
their permutation symmetry ,as $M_\nu^{S_4}$, $M_\nu^{S_6}$
respectively \footnote{We study texture $M_\nu^{S_1}$ and its
permutation symmetry as texture $M_\nu^{S_2}$ in the
non-perturbation method and find their results which are exactly
consistent with the experimental data. We also find that textures
$M_\nu^{S_3}$, $M_\nu^{S_5}$ and their permutation symmetry
texture which are ruled out\cite{200}.}.

We studied, the texture $M_\nu^{S_7}$, with $(\mu,\mu)$ and
$(\tau,\tau)$ vanishing elements, which has both magic and
$\mu-\tau $ symmetry. Therefore its mixing matrix is $U_{TBM}$,
with $\theta_{13}\neq0$, $\delta=0$, and $\theta_{23}=45^\circ$.
Hence, we consider $M_\nu^{S_7}$ with the attendance of a small
contribution as perturbation matrix by employing the perturbation
method~\footnote{As we know the mixing angle $\theta_{13}$ is
small compared to other two angles, $\theta_{12}$ and,
$\theta_{23}$, also the magnitude of the ratio of two neutrino
mass-squared differences is $R_\nu=\frac{\delta m^2}{\Delta
m^2}\simeq~10^{-2}$.}, which generates simultaneously a tiny
$\delta m^2$ and small parameters in the neutrino mixing
component, such as, $U_{13}$, ($\theta_{13}$ and $\delta$), and
provides minor amendments to $\theta_{23}$. Therefore, CP
violation is investigated.

We reproduce the texture $M_\nu^{S_7}$ as unperturbed mass matrix
in the flavor basis which has a structure in which
$\theta_{13}=0$, $\delta m^2=0$, $\theta_{23}=45^\circ$, and
$\Delta m^2\neq0$. We work on the diagonal mass matrix of the
reproduced $M_\nu^{S_7}$ as unperturbed mass matrix in the mass
basis which has degeneracy $m_1^{(0)}=m_2^{(0)}$.

Our perturbation mass matrix is symmetric and complex in the mass
basis, and thereby, is non-Hermitian matrix. We obtain
$\theta_{13}$, $\delta$ and, $\delta m^2$ that all could arise
from a perturbation. The perturbation also affects the atmospheric
mixing angle $\theta_{23}$. We have two different cases; in the
case I, $\Delta$ and $\textit{Re}(\alpha)$ are negative while in
the case II, both are positive.

We compare our results with the experimental data in each case; in
the case I, we could obtain the allowed range of our parameter
space, $\frac{m}{\Delta}$ and the complex elements of perturbation
mass matrix $\alpha$, and $\beta$; then we check the accuracy of
our work and our results by obtaining the magnitude of
$\theta_{23}$ by the obtained allowed range of our parameters. We
obtain $\theta_{23}\approx(41.81^\circ-43.96^\circ)$ which is in
complete agreement with the experimental data in Eq.\,(\ref{exp})
and shows the accuracy of our work. Moreover we predict the
neutrino mass ordering is inverted in this case.

We fail to obtain an acceptable range for $\beta$ in the case II
so this case is ruled out.

\section{Acknowledgments}
We would like to thank the research office of the Qazvin Branch,
Islamic Azad University.


\begin{thebibliography}{10}


\bibitem{mixing1}
J. Schechter and J. W. F. Valle, Neutrino masses in
$SU(2)\bigotimes U(1)$ theories, Phys.Rev. D22 (1980) 2227.

\bibitem{mixing2}
 H. Fritzsch and Z.Z. Xing, How to Describe Neutrino Mixing and CP Violation
 , Phys. Lett. B 517, 363 (2001) [hep-ph/0103242v2].

\bibitem{mixing3}
Particle Data Group, W.M. Yao et al., Review of Particle Physics ,
J. Phys. G 33, 1 (2006).

\bibitem{26} K. Abe et al. (T2K Collaboration), Precise Measurement of the
Neutrino Mixing Parameter $\theta_{23}$ from Muon Neutrino
Disappearance in an Off-Axis Beam, Phys. Rev. Lett. 112, 181801
(2014).

\bibitem{27} K. Abe et al. (T2K Collaboration), Observation of Electron
Neutrino Appearance in a Muon Neutrino Beam, Phys. Rev. Lett. 112,
061802 (2014).

\bibitem{28} J. K. Ahn et al. (RENO Collaboration), Observation of Reactor
Electron Antineutrino Disappearance in the RENO Experiment, Phys.
Rev. Lett. 108, 191802 (2012).

\bibitem{29} Y. Abe et al. (Double Chooz Collaboration), Reactor electron
antineutrino disappearance in the Double Chooz experiment, Phys.
Rev. D 86, 052008 (2012).

\bibitem{30} F. P. An et al. (Daya Bay Collaboration), Spectral
Measurement of Electron Antineutrino Oscillation Amplitude and
Frequency at Daya Bay, Phys. Rev. Lett. 112, 061801 (2014).

\bibitem{31} B. Z. Hu (Daya Bay Collaboration), New results from the Daya
Bay reactor neutrino experiment, arXiv:1402.6439.

\bibitem{32} F. Capozzi, G. L. Fogli, E. Lisi, A. Marrone, D. Montanino,
and A. Palazzo, Status of three-neutrino oscillation parameters,
circa 2013, Phys. Rev. D 89, 093018 (2014).

\bibitem{33} K. A. Olive, K. Nakamura, S. T. Petcov et al. (Particle Data
Group Collaboration), Review of particle physics, Chin. Phys. C
38, 090001 (2014).

\bibitem{TBM}
P. F. Harrison et al., Tri-Bimaximal Mixing and the Neutrino
Oscillation Data, Phys. Lett. B530, 167 (2002) [hep-ph/0202074v1].

\bibitem{25} P. F. Harrison, D. H. Perkins, and W. G. Scott, Tri-bimaximal
mixing and the neutrino oscillation data, Phys. Lett. B 530, 167
(2002).

\bibitem{256}P. H. Frampton , T. W.
Kephart and S. Mat-suzaki, Simplified renormalizable T model for
tribimaximal mixing and Cabibbo angle, Phys. Rev. D 78 , 073004
(2008) [hep-ph/0807.4713].

\bibitem{259}G. Altarelli and F. Feruglio,
Tri-Bimaximal Neutrino Mixing, A4 and the Modular Symmetry, Nucl.
Phys. B 741 , 215 (2006) [hep-ph/0512103].

\bibitem{2591}
B. Grinstein and M. Trott, [hep-ph/1203.4410]; S. F. King,
Parametrizing the lepton mixing matrix in terms of deviations from
tri-bimaximal mixing, Phys. Lett. B 659, 244 (2008)
[hep-ph/0710.0530]; S. Pakvasa, W. Rodejohann, T. Weiler, Unitary
Parametrization of Perturbations to Tribimaximal Neutrino Mixing,
Phys. Rev. Lett. 100, 111801 (2008); C. H. Albright, A. Dueck, W.
Rodejohann, Possible Alternatives to Tri-bimaximal Mixing, Eur.
Phys. J. C 70, 1099-1110 (2010) [hep-ph/1004.2798v1]; S. Boudjemaa
and S. F. King, Deviations from Tri-bimaximal Mixing: Charged
Lepton Corrections and Renormalization Group Running, Phys. Rev. D
79, 033001 (2009) [hep-ph/0808.2782]; S. Goswami, S. T. Petcov, S.
Ray and W. Rodejohann, Large Ue3 and Tri-bimaximal Mixing, Phys.
Rev. D 80, 053013 (2009) [hep-ph/0907.2869]; D. Meloni, F.
Plentinger and W. Winter, Perturbing exactly tri-bimaximal
neutrino mixings with charged lepton mass matrices, Phys. Lett. B
699, 354 (2011) [hep-ph/1012.1618]; Sumit K. Garg, Consistency of
perturbed Tribimaximal, Bimaximal and Democratic mixing with
Neutrino mixing data , Nucl. Phys. B931C (2018) 469-505; D.
Marzocca, S. T. Petcov, A. Romanino and M. Spinrath, Sizeable è13
from the Charged Lepton Sector in SU(5), (Tri-)Bimaximal Neutrino
Mixing and Dirac CP Violation, JHEP 1111, 009 (2011)
[hep-ph/1108.0614]; G. Altarelli and F. Feruglio, Tri-Bimaximal
Neutrino Mixing, A4 and the Modular Symmetry, Nucl. Phys. B 741,
215 (2006) [hep-ph/0512103]; F. Bazzocchi, S. Morisi and M.
Picariello, Embedding A4 into left-right flavor symmetry:
Tribimaximal neutrino mixing and fermion hierarchy, Phys. Lett. B
659, 628 (2008) [hep-ph/0710.2928]; E. Ma and D. Wegman, Nonzero
theta(13) for neutrino mixing in the context of A(4) symmetry,
Phys. Rev. Lett. 107, 061803 (2011) [hep-ph/1106.4269]; S. Gupta,
A. S. Joshipura and K. M. Patel, Minimal extension of
tri-bimaximal mixing and generalized Z2 X Z2 symmetries, Phys.
Rev. D 85, 031903 (2012) [hep-ph/1112.6113]; B. Adhikary, B.
Brahmachari, A. Ghosal, E. Ma and M. K. Parida, A4 symmetry and
prediction of Ue3 in a modified Altarelli-Feruglio model, Phys.
Lett. B 638, 345 (2006) [hep-ph/0603059]; E. Ma, Near Tribimaximal
Neutrino Mixing with Ä(27) Symmetry, Phys. Lett. B 660, 505 (2008)
[hep-ph/0709.0507]; F. Plentinger, G. Seidl and W. Winter, Group
space scan of flavor symmetries for nearly tribimaximal lepton
mixing, JHEP 0804, 077 (2008) [hep-ph/0802.1718]; N. Haba, R.
Takahashi, M. Tanimoto and K. Yoshioka, Tri-bimaximal Mixing from
Cascades, Phys. Rev. D 78, 113002 (2008) [hep-ph/0804.4055]; S.
-F. Ge, D. A. Dicus and W. W. Repko, Residual Symmetries for
Neutrino Mixing with a Large theta13 and Nearly Maximal deltaD,
Phys. Rev. Lett. 108, 041801 (2012) [hep-ph/1108.0964]; T. Araki
and Y. F. Li, Q6 flavor symmetry model for the extension of the
minimal standard model by three right-handed 15 sterile neutrinos,
Phys. Rev. D 85, 065016 (2012) [hep-ph/1112.5819].

\bibitem{ma}
E. Ma and G. Rajasekaran, Phys. Rev. D64, 113012 (2001).

\bibitem{ma1}
Ernest Ma, Aspects of the Tetrahedral Neutrino Mass Matrix,
Phys.Rev.D72:037301,(2005).

\bibitem{16} G. Altarelli and F. Feruglio, Discrete flavor symmetries and
models of neutrino mixing, Rev. Mod. Phys. 82, 2701 (2010).

\bibitem{17}
M. Hirsch, A. S. Joshipura, S. Kaneko, and J.W. F. Valle,
Predictive Flavour Symmetries of the Neutrino Mass Matrix, Phys.
Rev. Lett. 99, 151802 (2007).

\bibitem{18} G. Altarelli and F. Feruglio, Tri-bimaximal neutrino mixing,
A(4) and the modular symmetry, Nucl. Phys. B741, 215 (2006).

\bibitem{19} G. Altarelli and D. Meloni, A simplest A4 model for
tribimaximal neutrino mixing, J. Phys. G 36, 085005 (2009).

\bibitem{20} K. M. Parattu and A. Wingerter, Tribimaximal mixing from
small groups, Phys. Rev. D 84, 013011 (2011).

\bibitem{21} S. F. King and C. Luhn, A4 models of tri-bimaximal-reactor
mixing, J. High Energy Phys. 03 (2012) 036.

\bibitem{22} G. Altarelli, F. Feruglio, L. Merlo, and E. Stamou, Discrete
flavour groups, $\theta_{13}$ and lepton flavour violation, J.
High Energy Phys. 08 (2012) 021.

\bibitem{23} G. Altarelli, F. Feruglio, and L. Merlo, Tri-bimaximal
neutrino mixing and discrete flavour symmetries, Fortschr. Phys.
61, 507 (2013).

\bibitem{24} P. M. Ferreira, L. Lavoura, and P. O. Ludl, A new A4 model
for lepton mixing, Phys. Lett. B 726, 767 (2013).

\bibitem{38} J. Barry and W. Rodejohann, Deviations from tribimaximal
mixing due to the vacuum expectation value misalignment in $A_4$
models, Phys. Rev. D 81, 093002 (2010); 81, 119901(E) (2010).

\bibitem{39}
Y. H. Ahn and S. K. Kang, Non-zero $\theta_{13}$ and CP violation
in a model with $A_4$ flavor symmetry, Phys. Rev. D 86, 093003
(2012).

\bibitem{43} H. Ishimori and E. Ma, New simple A4 neutrino model for
nonzero $\theta_{13}$ and large $\delta_{CP}$, Phys. Rev. D 86,
045030 (2012).

\bibitem{44} E. Ma, A. Natale, and A. Rashed, Scotogenic A4 neutrino model
for nonzero $\theta_{13}$ and large $\delta_{CP}$, Int. J. Mod.
Phys. A 27, 1250134 (2012).

\bibitem{45} D. N. Dinh, N. A. Ky, P. Q. V.n, and N. T. H. Van, in 2nd
International Workshop on Theoretical and Computational Physics
(IWTCP-2), Ban-Ma-Thuat, July, 2014 (2014); A prediction of
$\delta_{CP}$ for a normal neutrino mass ordering in an extended
standard model with an $A_4$ flavour symmetry, J. Phys. Conf. Ser.
627, 012003 (2015).

\bibitem{46} D. N. Dinh, N. A. Ky, P. Q. Vãn, and N. T. H. Vân, A seesaw
scenario of an $A_4$ flavor symmetric standard model,
arXiv:1602.07437.

\bibitem{exp}
P. F. de Salas, D. V. Forero, S. Gariazzo, P. Martínez-Miravé, O.
Mena, C. A. Ternes, M. Tórtola, J. W. F. Valle, 2020 Global
reassessment of the neutrino oscillation picture,     J. High
Energ. Phys. 2021, 71 (2021).

\bibitem{200}
Razzaghi, N.; Rasouli, S.M.M.; Parada, P.; Moniz, P. Two-Zero
Textures Based on A4 Symmetry and Unimodular Mixing Matrix.
Symmetry 2022, 14, 2410.

\bibitem{pur39}
F. Vissani, J. High Energy Phys. 9811 (1998) 025; E.K.Akhmedov,
Phys. Lett. B 467 (1999) 95; M. Lindner, W. Rodejohann, J. High
Energy Phys. 0705 (2007) 089; D. Aristizabal Sierra, I. de
Medeiros Varzielas, E. Houet, Phys. Rev. D 87 (2013) 093009; T.
Araki, Prog. Theor. Exp. Phys. 2013 (2013) 103B02; M.-C. Chen, J.
Huang, K.T. Mahanthappa, A.M. Wijangco, J. High Energy Phys. 1310
(2013) 112; L.J. Hall, G.G. Ross, J. High Energy Phys. 1311 (2013)
091; Jiajun Liao, D. Marfatia, K. Whisnant, Phys. Rev. D 92,
073004 (2015) ; Sumit K. Garg, Nucl. Phys. B931C (2018) 469-505.

\bibitem{magic}
C.S. Lam, Magic Neutrino Mass Matrix and the
Bjorken-Harrison-Scott Parameterization, Phys.Lett. B640 (2006)
260-262[ arXiv:hep-ph/0606220v2].

\bibitem{kumar}
Radha Raman Gautam, Sanjeev Kumar, Zeros in the magic neutrino
mass matrix, PHYSICAL REVIEW D 94, 036004 (2016).

\bibitem{z8} S.Weinberg, Trans. New York Acad. Sci. 38, 185 (1977);
F.Wilczek and A. Zee, Phys. Lett. 70B, 418 (1977); H. Fritzsch,
Phys. Lett. 70B, 436 (1977); Phys. Lett. 73B, 317 (1978); Nucl.
Phys. B155, 18 (1979).

\bibitem{z9} H. Fritzsch and Z.Z. Xing, Prog. Part. Nucl. Phys. 45, 1
(2000); Z.Z. Xing, Int. J. Mod. Phys. A 19, 1 (2004);
arXiv:hep-ph/0406049.

\bibitem{z10} Paul H. Frampton, Sheldon L. Glashow and Danny Marfatia,
Phys. Lett. B 536, 79 (2002),hep-ph/0201008. 20

\bibitem{z11} Zhi-zhong Xing, Phys. Lett. B 530, 159 (2002),
hep-ph/0201151; H. Fritzsch, Z. Z. Xing, Phys. Lett. B 517 (2001)
363-368, arXiv: hep ph/0103242.

\bibitem{z12} Bipin R. Desai, D. P. Roy and Alexander R. Vaucher, Mod.
Phys. Lett. A 18, 1355 (2003), hep-ph/0209035; A. Merle, W.
Rodejohann, Phys. Rev. D 73, 073012 (2006), hep ph/0603111; S.
Dev, Sanjeev Kumar, S. Verma and S. Gupta, Nucl. Phys. B 784,
103-117 (2007), hep-ph/0611313; S. Dev, S. Kumar, S. Verma and S.
Gupta, Phys. Rev. D 76, 013002 (2007), hep-ph/0612102; M.
Randhawa, G. Ahuja, M. Gupta, Phys. Lett. B 643, 175-181 (2006),
hep- ph/0607074; G. Ahuja, S.Kumar, M. Randhawa, M. Gupta, S. Dev,
Phys. Rev. D 76, 013006 (2007), hep-ph/0703005; S. Kumar, Phys.
Rev. D 84, 077301 (2011), arXiv: 1108.2137 [hep-ph]; P. O. Ludl,
S. Morisi, E. Peinado, Nucl. Phys. B 857, 411 (2012), arXiv:
1109.3393 [hep-ph]; Manmohan Gupta, Gul- sheen Ahuja, Int. J. Mod.
Phys. A, 27, 1230033 (2012), arXiv:1302.4823 [hep- ph]; D.Meloni,
G. Blankenburg, Nucl. Phys. B 867, 749 (2013), arXiv:1204.2706
[hep-ph]; W. Grimus, P. O. Ludl, J. Phys. G40, 055003 (2013),
arXiv:1208.4515 [hep-ph]; S. Sharma, P. Fakay, G. Ahuja and M.
Gupta arXiv: 1402.1598 [hep- ph]; P. O. Ludl, W. Grimus,
arXiv:1406.3546v1 [hep-ph]; H. Fritzsch, Zhi-zhong Xing, S. Zhou,
JHEP 1109, 083 (2011), arXiv: 1108.4534 [hep-ph]; Madan Singh,
Gulsheen Ahuja, Manmohan Gupta, Prog. Theor. Exp. Phys. (PTEP)
2016 (12): 123B 08, arXiv: 1603.08083 [hep-ph]; Madan Singh, Adv.
High Energy Phys. 2018(2018) 2863184, arXiv: 1803.10735[hep-ph].

\bibitem{z13} C. Hagedorn and W. Rodejohann, JHEP 0507, 034 (2005).

\bibitem{z14}
X. Liu, S. Zhou, Int. J. Mod. Phys. A 28 (2013) 1350040.

\bibitem{z15}
E. I. Lashin and N. Chamoun, Phys. Rev. D85, 113011 (2012), arXiv:
1108.4010 [hep-ph].


\bibitem{permutation}
H. Fritzsch, Z.-z. Xing, and S. Zhou, J. High Energy Phys. 09
(2011) 083.

\bibitem{meFL3}
Razzaghi, N.; Rasouli, S.M.M.; Parada, P.; Moniz, P. Generating CP
Violation from a Modified Fridberg-Lee Model. Universe 2022, 8,
448.

\bibitem{planck}
Aghanim, N.; Akrami, Y.; Ashdown, M.; Aumont, J.; Baccigalupi, C.;
Ballardini, M.; Banday, A.J.; Barreiro, R.B.; Bartolo, N.; Basak,
S.; et al. Planck 2018 results-VI. Cosmological parameters.
Astron. Astrophys. 2020, 641, A6.
























...






\end{thebibliography}
\end{document}